\documentclass[manuscript]{aastex}
\usepackage{graphicx}

\shorttitle{A new GPU-accelerated hydrodynamical code for numerical simulation of interacting galaxies}
\shortauthors{Kulikov I.}

\begin{document}

\title{GPUPEGAS: a new GPU-accelerated hydrodynamical code for numerical simulation of interacting galaxies}

\author{Igor Kulikov\altaffilmark{1}}
\affil{Institute of Computational Mathematics and Mathematical Geophysics SB RAS, \\ Russia, Novosibirsk, 630090}

\altaffiltext{1}{Ph. D., Junior research scientist, Department of parallel algorithms of large-scale problems solving,
Institute of Computational Mathematics and Mathematical Geophysics SB RAS, 630090, Novosibirsk, Russia,
Associate Professor, Novosibirsk State Technical University, 630092, Novosibirsk, Russia, kulikov@ssd.sscc.ru}

\begin{abstract}
In this paper a new scalable hydrodynamic code GPUPEGAS (GPU-accelerated PErformance Gas Astrophysic Simulation) 
for simulation of interacting galaxies is proposed. The code is based on combination of
Godunov method as well as on the original implementation of FlIC method, specially adapted for GPU-implementation.
Fast Fourier Transform is used for Poisson equation solution in GPUPEGAS.
Software implementation of the above methods was tested on classical gas dynamics problems, new Aksenov's test and 
classical gravitational gas dynamics problems. Collisionless hydrodynamic approach was used for modelling of stars and dark matter. 
The scalability of GPUPEGAS computational accelerators is shown.
\end{abstract}


\section{Introduction}
The movement of galaxies in dense clusters turns the collisions of galaxies into an important 
evolutionary factor \citep{Tutukov_2006,Tutukov_2011}. Numerical simulation plays a major role 
in studying of these processes. Due to extremely high growth of supercomputers performance the 
new astrophysical models and codes needs to be developed for detailed simulation of different 
physical effects in astrophysics.

During the last 15 years, from the wide range of the hydrodynamical methods two main approaches 
are used for non-stationary astrophysical problems solution. They are the Lagrangian SPH 
method \citep{Gingold_1977,Luci_1977} (Smoothed Particle Hydrodynamics) and  Eulerian methods within 
adaptive meshes, or AMR \citep{Norman_2005} (Adaptive Mesh Refinement). During the last 5 years a lot 
of combined codes (using both Lagrangian and Eulerian approaches) appeared.

The main problem of SPH codes is the search of neighbors for a partcle and the computation of gravitational
interaction between particles. In order to solve these problems a lot of algorithms were developed: 
particle-particle/particle-mesh or P$^{3}$M method \citep{Hockney_1981}, adaptation of P$^{3}$M
method with using of hierarchy of computational mesh AP$^{3}$M \citep{Couchman_1991}, tree code \citep{Barnes_1986}, 
combination of tree code and particle-mesh method - Tree-PM method \citep{Dubinski_2004}. 
There are some methods using for solving of Poisson equation: the conjugate gradient method, 
the fast Fourier transform method, the method of successive over-relaxation and Fedorenko method (The Multigrid
method) \citep{Fedorenko_1961}.

For numerical solution of gas dynamics problems the Godunov method is widely used\citep{Godunov_1959}, that's main 
structural element is a Riemann problem. Different algorithms of Riemann problem solution have produced a wide class of mathematical methods \citep{Kulikovsky_2001,Toro_1999}. The main methods are Courant - Isakson - Reese 
method \citep{Courant_1952} and Roe method \citep{Roe_1997}. These methods are based on linearized hyperbolic systems 
of equations \citep{Engquist_1981} where the solution of Riemann problem is constructed only of Riemann waves.
The main approach of wave velocity evaluation is a double wave Harten-Lax-Van Leer(HLL) scheme \citep{Harten_1983} 
and his modification HLLE-method \citep{Einfeld_1988}, HLLC \citep{Batten_1997}.
Some schemes, based on Godunov method, such as upwind second order MUSCL \citep{VanLeer_1979} and TVD 
schemes \citep{Jin_1995}, third order piecewise parabolic method PPM \citep{Collela_1984} were developed. 
Still is is not clear how to determine the order of accuracy of a scheme in the case of a discontinuous 
solution as it is stated in \citep{Godunov_2011}.

The most well-known SPH codes are Hydra \citep{Couchman_9603116},
Gasoline \citep{Wadsley_2004}, GrapeSPH \citep{Matthias_1996}, GADGET \citep{Springel_2005}.
The most well-known mesh codes (in some case with adaptive mesh refinement) are 
NIRVANA \citep{Ziegler_2005}, FLASH \citep{Mignone_2005}, ZEUS-MP \citep{Hayes_2006}, 
ENZO \citep{Norman_2005}, RAMSES \citep{Teyssier_2002}, ART \citep{Kravtsov_2002}, 
Athena \citep{Stone_2008}, Pencil Code \citep{Brandenburg_0109497}, Heracles \citep{Gonzalez_2007},
Orion \citep{Krumholz_2007}, Pluto \citep{Mignone_2007}, CASTRO \citep{Almgren_2010}, 
GAMER \citep{Schive_2010}. BETHE-Hydro \citep{Murphy_2008}, AREPO \citep{Springel_2009}, 
CHIMERA \citep{Bruenn_2009} and PEGAS \citep{Vshivkov_2011} (designed by the author of the present paper) 
codes are based on mixed Lagrangian and Eulerian approaches.
The large number of existing astrophysical codes means that there is no perfect code suitable for all cases. 
In such a way, new code development as well as modification of an existing codes is still necessary.
In spite of the development of PETAFLOPs astrophysics codes such as  PetaGADGET \citep{Feng_2011}, Enzo-P, PetaART, it is necessary to note some scalability restrictions of both AMR and SPH methods \citep{Ferrari_2010,Straalen_2009}. The main properties of codes are given in the table \ref{PropTable}.

The present paper contain description a new collisionless component of galaxies on 
basis ''Collisionless hydrodynamic'' model. The first time, the model for astrophysical problems was 
used in \citep{Mitchell_2012}. GPUPEGAS is a first code, which contains full Boltzmann moment equations 
with complete the symmetric velocity dispersion tensor. The purpose of this paper is a description of the 
details of the numerical method as well as the peculiarities of the implementation of the method for 
hydrid supercomputers equipped with GPU-accelerated. 

\section{Numerical Method Description}
We should use a two-phase approach for modelling of interacting galaxies: hydrodynamic component \citep{Vshivkov_2011}
and collisionless component \citep{Mitchell_2012} for description of stars and dark matter. 

We should use for hydrodynamic component is the 3D model of self-gravitating gas dynamics in Cartesian coordinate system.
$$
\frac{\partial \rho}{\partial t} + \frac{\partial \rho v_{k}}{\partial x_{k}} = 0,
$$
$$
\frac{\partial \rho v_{i}}{\partial t} + \frac{\partial \rho v_{i} v_{k}}{\partial x_{k}}
= - \frac{\partial p}{\partial x_{i}} - \rho \frac{\partial (\Phi + \Phi_{0})}{\partial x_{i}}
$$
$$
\frac{\partial \rho E}{\partial t} + \frac{\partial \rho E v_{k}}{\partial x_{k}} =
- \frac{\partial p v_{k}}{\partial x_{k}} - \rho v_{k} \frac{\partial (\Phi + \Phi_{0})}{\partial x_{k}} - q,
$$
$$
\frac{\partial \rho \epsilon}{\partial t} + \frac{\partial \rho \epsilon v_{k}}{\partial x_{k}} =
- (\gamma - 1)\rho \epsilon \frac{\partial v_{k}}{\partial x_{k}} - q,
$$
$$
\Delta \Phi = 4\pi\rho,
$$
$$
p = (\gamma - 1)\rho \epsilon,
$$
$$
\rho E = \rho \epsilon + \frac{\rho v_{k}^{2}}{2},
$$
where $p$ is the pressure, $\rho$ is the density, $\vec{v} = (v_{x},v_{y},v_{z})$  is the velocity vector,
$\rho E$ is the density of total energy, $\Phi$ is the gravitational potential of the gas itself,
$\Phi_{0}$ is the contribution of the dark matter and stars to the gravitational potential,
$\epsilon$ - the inner energy, $q$ - cooling function \citep{Sutherland_1993}.

The dynamics of collisionless component are described by the collisionless Boltzmann equation 
for the distribution function of particles $f(x,t,w)$ in the 6D position($x$) -- velocity($w$) phase space:
$$
\frac{\partial f}{\partial t} + w_{k} \frac{\partial f}{\partial x_{k}} + g_{k} \frac{\partial f}{\partial w_{k}} = 0.
$$
The first moment of the collisionless Boltzmann equation are:
$$
n = \int m f d^{3}w,
$$ 
$$
n \vec{u} = \int m f w d^{3}w,
$$
$$
\Pi_{ij} = \int m f (w_{i} - u_{i}) (w_{j} - u_{j}) d^{3}w = \Pi_{ji},
$$ 
$$
n E_{ij} = \Pi_{ij} + n u_{i} u_{j},
$$
where $\Pi_{ij}$ is the symmetric velocity dispersions tensor, $n$ is the density, $\vec{u} = (u_{x},u_{y},u_{z})$  is the velocity vector, $n E_{ij}$ is the density of total energy, $\Phi$ is the gravitational potential of the gas itself,
$\Phi_{0}$ is the contribution of the dark matter and stars to the gravitational potential,
$m$ is the particles mass.

We should use for collisionless component is the 3D model of Boltzmann moment equations in Cartesian coordinate system.
$$
\frac{\partial n}{\partial t} + \frac{\partial n u_{k}}{\partial x_{k}} = 0,
$$
$$
\frac{\partial n u_{i}}{\partial t} + \frac{\partial n u_{i} u_{k}}{\partial x_{k}} = 
- \frac{\partial \Pi_{ik}}{\partial x_{k}} - n \frac{\partial (\Phi + \Phi_{0})}{\partial x_{i}}
$$
$$
\frac{\partial n E_{ij}}{\partial t} + \frac{\partial n E_{ij} u_{k}}{\partial x_{k}} =
- \frac{\partial ( \Pi{jk} u_{i} +  \Pi{ik} u_{j} )}{\partial x_{k}}
- n u_{i} \frac{\partial (\Phi + \Phi_{0})}{\partial x_{j}} - n u_{j} \frac{\partial (\Phi + \Phi_{0})}{\partial x_{i}}
$$
$$
\frac{\partial \Pi_{ij}}{\partial t} + \frac{\partial \Pi_{ij} u_{k}}{\partial x_{k}} =
- \Pi_{jk} \frac{\partial u_{i}}{\partial x_{k}} - \Pi_{ik} \frac{\partial u_{j}}{\partial x_{k}},
$$
$$
\Delta \Phi_{0} = 4 \pi n,
$$

The main characteristic parameters are: $L = 10000$ parsec, $M_{0} = 10^{11} M_{\odot}$, 
$G = 6.67 \cdot 10^{-11}$ N m$^{2}$/kg, $q = 2 \cdot 10^{-24}$ kg/sec $^{3}$ m.
Let us introduce a uniform grid in the 3D computation domain. The cells of the grid are:
$x_{i}=ih_{x}$, $i=1,..,I_{max}$, $y_{k}=kh_{y}$, $k=1,..,K_{max}$, $z_{l}=lh_{z}$, $l=1,..,L_{max}$,
where $h_{x}$, $h_{y}$, $h_{z}$  are the mesh steps, $I_{max}$, $K_{max}$, $L_{max}$ are the numbers of the mesh cells
along the directions $x$, $y$, $z$: $h_{x}=x_{max}/I_{max}$, $h_{y}=y_{max}/K_{max}$, $h_{z}=z_{max}/L_{max}$.
The method for the solution of gas dynamics equation is based on the Fluids-In-Cells and Godunov method \citep{Vshivkov_2011},
which showed good advantage for astrophysical problems \citep{Tutukov_2011}.

\subsection{The gas dynamics equation solution technique}
The solution of the gas dynamics equations system is performed in two stages.
During the first (Eulerian) stage, the equations system describes the the change of gas values as 
the result of pressure, gravity and cooling. The operator approach is used for the elimination of the mesh effect
\cite{Vshivkov_2006}. The pressure $P$ and velocity $V$ values at all cells boundaries are 
exact solution of the linearized Eulerean stage equation system without potential and cooling function.

Let us consider 1D gas dynamics equations in Cartesian coordinate system. 
$$
\frac{\partial \rho}{\partial t} + [v \frac{\partial \rho}{\partial x}] + \rho \frac{\partial v}{\partial x} = 0,
$$
$$
\frac{\partial v}{\partial t} + [v \frac{\partial v}{\partial x}] + \frac{1}{\rho} \frac{\partial p}{\partial x} = 0,
$$
$$
\frac{\partial p}{\partial t} + [v \frac{\partial p}{\partial x}] + \gamma p \frac{\partial v}{\partial x} = 0.
$$
We can rule out advective terms and consider 1D gas dynamics equations on Eulerian stage:
$$
\frac{\partial}{\partial t}
\left(%
\begin{array}{c}
  \rho \\ 
  v \\ 
  p \\
\end{array}%
\right) +
\left(%
\begin{array}{ccc}
  0 & \rho & 0 \\ 
  0 & 0 & \rho^{-1} \\ 
  0 & \gamma p & 0 \\
\end{array}%
\right)
\frac{\partial}{\partial x}
\left(%
\begin{array}{c}
  \rho \\ 
  v \\ 
  p \\
\end{array}%
\right) = 0
$$  
The eigenvalues of this matrix is:
$\lambda_{1} = 0, \lambda_{2} = \sqrt{\frac{\gamma p}{\rho}}, \lambda_{3} = -\sqrt{\frac{\gamma p}{\rho}}$.
We should rule out first column and first row and consider equations:
$$
\frac{\partial q}{\partial t} + B \frac{\partial q}{\partial x} = 0,
$$
where $q = (v,p)$, $B= R \Lambda L$, $R$ -- matrix of right eigenvectors, $L$ -- matrix of left eigenvectors,
$\Lambda$ -- diagonal matrix of eigenvalues, $LR = I$. We should make the change $s = Lq$ and 
consider independent equations:
$$
\frac{\partial s}{\partial t} + \Lambda \frac{\partial s}{\partial x} = 0.
$$
This system of equations has the exact solution at each cell boundaries, depending on the sign of the eigenvalues.
We should make the inverse change $q = R s$ and $q$ is the exact solution of equations on the Eulerian stage.

The eigenvalues and eigenvectors of 1D gas dynamics equations on Eulerian stage is:
$$
\lambda_{1} = \sqrt{\frac{\gamma p}{\rho}},
$$
$$
r_{1} = \left( \frac{1}{\sqrt{\gamma p \rho}}, \frac{\rho \sqrt{\gamma p / \rho}}{\sqrt{\gamma p \rho}}  \right),
$$
$$
l_{1} = \left( \frac{1}{\sqrt{\gamma p \rho}}, \frac{1}{\sqrt{\gamma p \rho}} \right).
$$
$$
\lambda_{2} = -\sqrt{\frac{\gamma p}{\rho}},
$$
$$
r_{2} = \left( \frac{1}{\sqrt{\gamma p \rho}}, -\frac{\rho \sqrt{\gamma p / \rho}}{\sqrt{\gamma p \rho}}  \right),
$$
$$
l_{2} = \left( \frac{\rho \sqrt{\gamma p / \rho}}{\sqrt{\gamma p \rho}}, -\frac{\rho \sqrt{\gamma p / \rho}}{\sqrt{\gamma p \rho}}  \right).
$$
This system is linearly hyperbolic and it has the following analytic solution:
$$
V = \frac{v_{L} + v_{R}}{2} + \frac{p_{L} - p_{R}}{2} \sqrt{\frac{\rho_{L}+\rho_{R}}{\rho_{L}\rho_{R} \gamma (p_{L} + p_{R})}}
$$
$$
P = \frac{p_{L} + p_{R}}{2} + \frac{v_{L} - v_{R}}{2} \sqrt{\frac{\rho_{L}\rho_{R} \gamma (p_{L} + p_{R})}{\rho_{L}+\rho_{R}}}
$$
where $f_{L}, f_{R}$ corresponds to the values of a function left and right at the cells boundaries. 
This values are used in Eulerean stage scheme.

\begin{figure}[ht]
\centering
\includegraphics[bb = 0 0 674 506, width=0.6\linewidth]{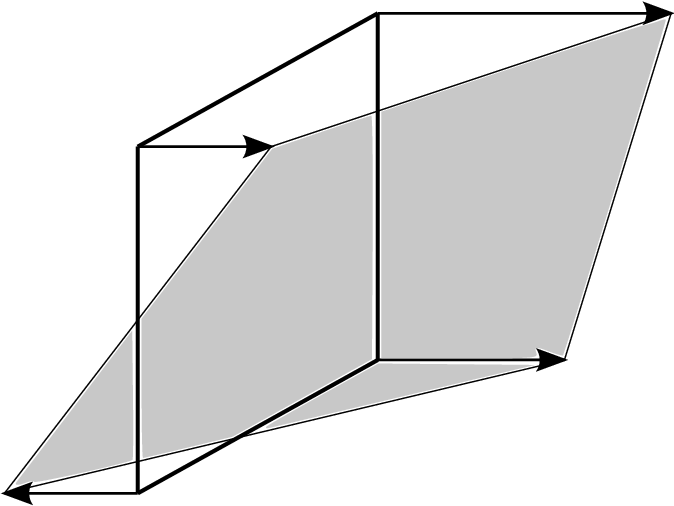}
\caption{Stream gasdynamic quantities across the boundary is defined by the rule of the total deformation of the cell}
\label{lagrange}
\end{figure}

During the second (Lagrangian) stage, the equations system contain divergent equations of the following type:
$$
\frac{\partial f}{\partial t} + div(f \vec{v}) = 0,
$$
The Lagrangian stage describes the advective transportation process of all gas quantities $f$. 
The initial version of numerical method involved the computation of the contributions of the gas quantities to
adjacent cells \cite{Vshivkov_2007}. The computation was based on the scheme velocity . But this approach 
is not suitable for computation accelerators (Nvidia Tesla or Intel Xeon Phi etc.). In order to show it 
let us consider the solution of the above equation in 1D form:
$$
\frac{f_{ikl}^{n+1}-f_{ikl}^{n}}{\tau} + \frac{F_{i+1/2,kl}^{n+1/2} - F_{i-1/2,kl}^{n+1/2}}{h} = 0
$$
where $F_{i+1/2,kl}^{n+1/2}$ is defined as follows pic. \ref{lagrange}:
$$
F_{i+1/2,kl}^{n+1/2} = \frac{\sum v_{i+1/2,k \pm 1,l \pm 1} f_{ikl}^{+}}{4}
$$
which is demonstrated by pic. \ref{lagrange}.

\subsection{The Boltzmann moment equations solution technique}
We can rule out advective terms on the 1D Boltzmann moment equations and consider 
six (instead of 10 equations) system equations on Eulerian stage:
$$
\frac{\partial}{\partial t}
\left(%
\begin{array}{c}
  u_{x} \\ 
  u_{y} \\ 
  u_{z} \\ 
  \Pi_{xx} \\ 
  \Pi_{xy} \\ 
  \Pi_{xz} \\ 
\end{array}%
\right) +
\left(%
\begin{array}{cccccc}
  0 & 0 & 0 & \rho^{-1} & 0 & 0 \\ 
  0 & 0 & 0 & 0 & \rho^{-1} & 0 \\ 
  0 & 0 & 0 & 0 & 0 & \rho^{-1} \\ 
  3\Pi_{xx} & 0 & 0 & 0 & 0 & 0 \\
  2\Pi_{xy} & \Pi_{xx} & 0 & 0 & 0 & 0 \\
  2\Pi_{xz} & 0 & \Pi_{xx} & 0 & 0 & 0 \\      
\end{array}%
\right)
\frac{\partial}{\partial x}
\left(%
\begin{array}{c}
  u_{x} \\ 
  u_{y} \\ 
  u_{z} \\ 
  \Pi_{xx} \\ 
  \Pi_{xy} \\ 
  \Pi_{xz} \\ 
\end{array}%
\right) = 0
$$
We should repeat the approach described in the previous paragraph.
The eigenvalues and eigenvectors of 1D Boltzmann moment equations on Eulerian stage is:
$$
\lambda_{1} = \sqrt{\frac{3\Pi_{xx}}{\rho}},
$$
$$
r_{1} = \left(   
\frac{\sqrt{\Pi_{xx}}}{\Pi_{xz}\sqrt{3 \rho}}, \frac{\Pi_{xy}\sqrt{3}}{\Pi_{xz}\sqrt{\rho \Pi_{xx}}},
\frac{1}{\sqrt{3 \rho \Pi_{xx}}}, \frac{\Pi_{xx}}{\Pi_{xz}}, \frac{\Pi_{xy}}{\Pi_{xz}}, 1
\right),
$$
$$
l_{1} = \left(   
\frac{\rho \Pi_{xz} \sqrt{3}}{2 \sqrt{\rho \Pi_{xx}}}, 0, 0, \frac{\Pi_{xz}}{2 \Pi_{xx}}, 0, 0
\right),
$$

$$
\lambda_{2} = -\sqrt{\frac{3\Pi_{xx}}{\rho}},
$$
$$
r_{2} = \left(   
-\frac{\sqrt{\Pi_{xx}}}{\Pi_{xz}\sqrt{3 \rho}}, -\frac{\Pi_{xy}\sqrt{3}}{\Pi_{xz}\sqrt{\rho \Pi_{xx}}},
-\frac{1}{\sqrt{3 \rho \Pi_{xx}}}, \frac{\Pi_{xx}}{\Pi_{xz}}, \frac{\Pi_{xy}}{\Pi_{xz}}, 1
\right),
$$
$$
l_{2} = \left(   
-\frac{\rho \Pi_{xz} \sqrt{3}}{2 \sqrt{\rho \Pi_{xx}}}, 0, 0, \frac{\Pi_{xz}}{2 \Pi_{xx}}, 0, 0
\right),
$$

$$
\lambda_{3} = \sqrt{\frac{\Pi_{xx}}{\rho}},
$$
$$
r_{3} = \left(   
0, 0, \frac{1}{\sqrt{\rho \Pi_{xx}}}, 0, 0, 1
\right),
$$
$$
l_{3} = \left(   
-\frac{\Pi_{xz} \sqrt{\rho}}{2 \sqrt{\Pi_{xx}}}, 0, \frac{\sqrt{\rho \Pi_{xx}}}{2}, -\frac{\Pi_{xz}}{2 \Pi_{xx}}, 0, \frac{1}{2}
\right),
$$

$$
\lambda_{4} = \sqrt{\frac{\Pi_{xx}}{\rho}},
$$
$$
r_{4} = \left(   
0, \frac{1}{\sqrt{\rho \Pi_{xx}}}, 0, 0, 1, 0
\right),
$$
$$
l_{4} = \left(   
-\frac{\Pi_{xy} \sqrt{\rho}}{2 \sqrt{\Pi{xx}}}, \frac{\sqrt{\rho \Pi_{xx}}}{2}, 0, -\frac{\Pi_{xy}}{2 \Pi_{xx}}, \frac{1}{2}, 0
\right),
$$

$$
\lambda_{5} = -\sqrt{\frac{\Pi_{xx}}{\rho}},
$$
$$
r_{5} = \left(   
0, -\frac{1}{\sqrt{\rho \Pi_{xx}}}, 0, 0, 1, 0
\right),
$$
$$
l_{5} = \left(   
-\frac{\Pi_{xy} \sqrt{\rho}}{2 \sqrt{\Pi_{xx}}}, -\frac{\sqrt{\rho \Pi_{xx}}}{2}, 0, -\frac{\Pi_{xy}}{2 \Pi_{xx}}, \frac{1}{2}, 0
\right),
$$

$$
\lambda_{6} = -\sqrt{\frac{\Pi_{xx}}{\rho}},
$$
$$
r_{6} = \left(   
0, 0, -\frac{1}{\sqrt{\rho \Pi_{xx}}}, 0, 0, 1
\right),
$$
$$
l_{6} = \left(   
\frac{\Pi_{xz} \sqrt{\rho}}{2 \sqrt{\Pi_{xx}}}, 0, -\frac{\sqrt{\rho \Pi_{xx}}}{2}, -\frac{\Pi_{xz}}{2 \Pi_{xx}}, 0, \frac{1}{2}
\right).
$$

This system is linearly hyperbolic and it has the following analytic solution:
$$
u_{x} = \frac{u_{x}^{L} + u_{x}^{R}}{2} + \frac{\Pi_{xx}^{L} - \Pi_{xx}^{R}}{2} 
\left[ \frac{1}{\sqrt{3 \rho \Pi_{xx}}}  \right],
$$
$$
u_{y} = \frac{u_{y}^{L} + u_{y}^{R}}{2} + \frac{\Pi_{xy}^{L} - \Pi_{xy}^{R}}{2} 
\left[ \frac{1}{\sqrt{\rho \Pi_{xx}}}  \right] + \frac{\Pi_{xx}^{L} - \Pi_{xx}^{R}}{2} 
\left[ \frac{\Pi_{xy} (1-\sqrt{3}) }{\Pi_{xx} \sqrt{3 \rho \Pi_{xx}}}  \right],
$$
$$
u_{z} = \frac{u_{z}^{L} + u_{z}^{R}}{2} + \frac{\Pi_{xz}^{L} - \Pi_{xz}^{R}}{2} 
\left[ \frac{1}{\sqrt{\rho \Pi_{xx}}}  \right] + \frac{\Pi_{xx}^{L} - \Pi_{xx}^{R}}{2} 
\left[ \frac{\Pi_{xz} (1-\sqrt{3}) }{\Pi_{xx} \sqrt{3 \rho \Pi_{xx}}}  \right],
$$
$$
\Pi_{xx} = \frac{\Pi_{xx}^{L} + \Pi_{xx}^{R}}{2} + \frac{u_{x}^{L} - u_{x}^{R}}{2}
\left[ \sqrt{3 \rho \Pi_{xx}} \right],
$$
$$
\Pi_{xy} = \frac{\Pi_{xy}^{L} + \Pi_{xy}^{R}}{2} + \frac{u_{y}^{L} - u_{y}^{R}}{2} \left[ \sqrt{\rho \Pi_{xx}} \right] +
\frac{u_{x}^{L} - u_{x}^{R}}{2} \left[ \frac{\Pi_{xy} \rho (\sqrt{3} - 1) }{\sqrt{\rho \Pi_{xx}}}  \right],
$$
$$
\Pi_{xz} = \frac{\Pi_{xz}^{L} + \Pi_{xz}^{R}}{2} + \frac{u_{z}^{L} - u_{z}^{R}}{2} \left[ \sqrt{\rho \Pi_{xx}} \right] +
\frac{u_{x}^{L} - u_{x}^{R}}{2} \left[ \frac{\Pi_{xz} \rho (\sqrt{3} - 1) }{\sqrt{\rho \Pi_{xx}}}  \right].
$$
where $f_{L}, f_{R}$ corresponds to the values of a function left and right at the cells boundaries. 
Parameters in square the parentheses should be  averaged. This values are used in Eulerean stage scheme 
on The Boltzmann moment equations.

\subsection{The Poisson equation solution}
The Poisson equation solution for the gravitational potential 
$$
\Delta (\Phi + \Phi_{0}) = 4 \pi (\rho + n),
$$
is based on 27 points stencil. Fourier transform method is used to solve the Poisson equation. 
The Poisson equation solution scheme in Fourier space is:
$$
(\Phi+\Phi_{0})_{jmn} = \frac{4 \pi h^2 (\rho + n)_{jmn}}
{6(1-(1-\frac{2}{3}sin^{2}(\frac{\pi j}{I}))(1-\frac{2}{3}sin^{2}(\frac{\pi m}{K}))(1-\frac{2}{3}sin^{2}(\frac{\pi n}{L})))}
$$
The Fast Fourier Transform procedure is used for the direct and inverse transform.

\subsection{The Correctness checking}
On each step at time was used correctness checking:
$$
\kappa_{gas} = \int | \rho E - \rho \epsilon - \rho v^{2}/2 | dx
$$
$$
\kappa_{bme} = \int | n E_{ij} - \Pi_{ij} - n u_{i} u_{j} | dx
$$
The energy balance procedure on each time step described in \cite{Vshivkov_2011_2}.

\section{Parallel implementation}
\begin{figure}[ht]
\centering
\includegraphics[bb = 0 0 558 373, width=0.8\linewidth]{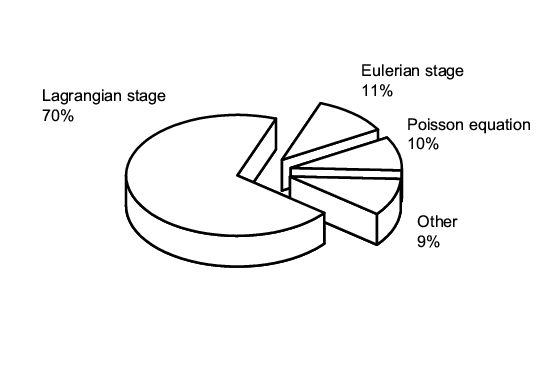}
\caption{The portion of each stage in the total computation time}
\label{percent}
\end{figure}

The necessity of three-dimensional simulation and the unsteady nature of the problem impose hard 
requirements for methods of solution. The rapid development of computer technology in recent time 
allowed to perform intensive computations and obtain physically correct results by means of the three-dimensional codes. 
Using of supercomputers gives the possibility to use large data volumes, improve the computation accuracy and 
also to accelerate computations. The main problem within the astrophysical code development is the efficient 
solution of gas dynamics equations and the Boltzmann moment equations since it takes up to 90\% of
computation time (fig. \ref{percent}).

\begin{figure}[ht]
\centering
\includegraphics[bb = 0 0 1416 1163, width=0.6\linewidth]{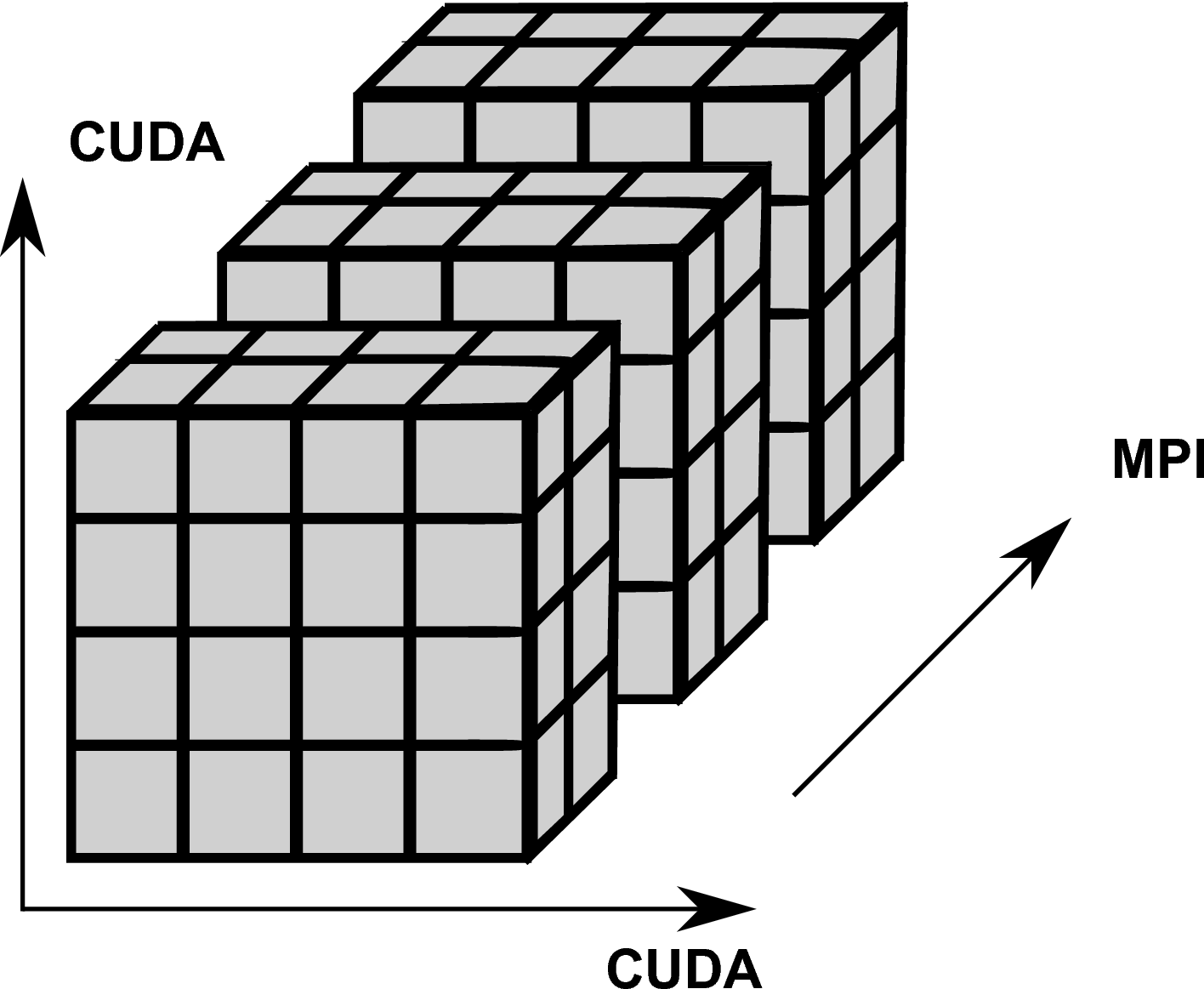}
\caption{Domain decomposition for the solution of the hydrodynamic equations}
\label{decomposition}
\end{figure}

The basis of the parallel implementation of  hydrodynamical solver is a three-dimensional domain decomposition.
There are MPI decomposition along the one coordinate, along another two coordinates CUDA technology are used  \ref{decomposition}. Three dimensional parallel Fast Fourier Transform is performed by the subroutine 
from the freeware FFTW library. In the future, such a Poisson solver will be ported to GPU accelerators.

Parallel implementation is related to the topology and architecture of a hybrid supercomputer NKS-G6 of 
Siberian Supercomputer Center ICMMG SB RAS. In basis of the modification of the numerical method for 
solving the hydrodynamic equations at every stage irrespective computing the fluxes through each cell.
Is performed the one-layer overlapping of the boundary points of the neighboring subdomains.
In the future, such a modification of the method will be ported to accelerators Intel Xeon Phi.

\begin{figure}[ht]
\centering
\includegraphics[bb = 0 0 486 399, width=0.45\linewidth]{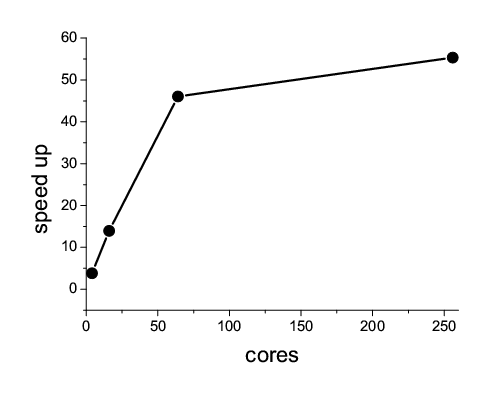}
\includegraphics[bb = 0 0 486 399, width=0.45\linewidth]{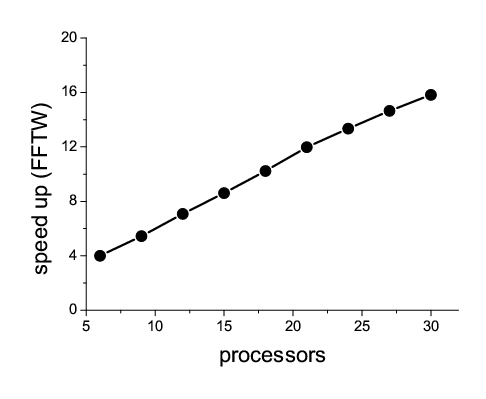}
\includegraphics[bb = 0 0 501 388, width=0.45\linewidth]{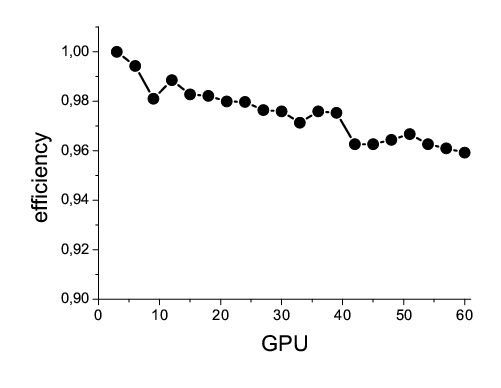} 
\caption{Speed up of gas dynamics equations and the Boltzmann moment equations on one GPU (first).
Speed up of the solver of Poisson equation, depending on the used cores (second).
Efficiency the parallel implementation of gas dynamics equations and the Boltzmann moment equations on the used GPU (third).}
\label{efficiency}
\end{figure}

In the case of a hybrid implementation is necessary to define two concepts of scalability:
\begin{itemize}
\item The strong scalability -- reduction of the computation time of one step of d the same problem when more devices are used.
\item The weak scalability -- a saving of computation time of one step  and the same amount of tasks with increasing the number of devices at the same time.
\end{itemize}
The results of the efficiency of program implementation are shown in the figure \ref{efficiency}.

\section{Testing of the implementation}
A GPUPEGAS code was verified on: 4 problems of shock tube (3 the Godunov tests for gas dynamics equations and 
one test for the Boltzmann moment equations); a new the Aksenov test; Kelvin-Helmholtz and Rayleigh-Taylor Instability tests; 
Author's cloud collision test; the collapse of rotating molecular cloud test.

\subsection{The Godunov tests}
\begin{figure}[ht]
\centering
\includegraphics[bb = 0 0 528 359, width=0.3\linewidth]{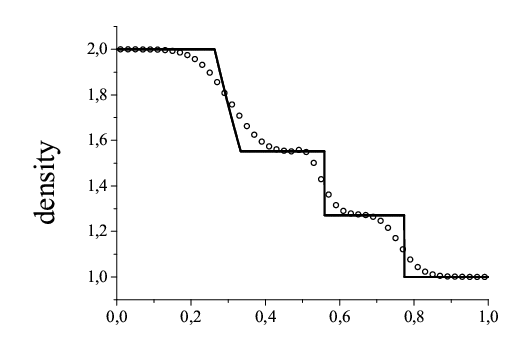}
\includegraphics[bb = 0 0 528 359, width=0.3\linewidth]{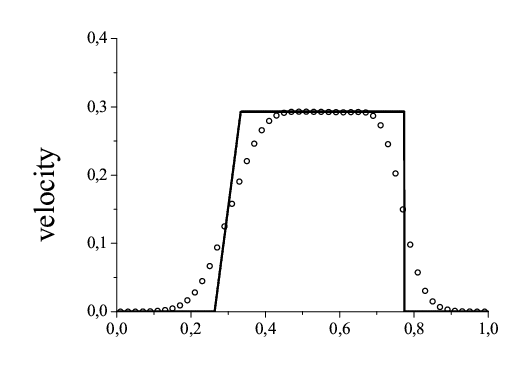}
\includegraphics[bb = 0 0 528 359, width=0.3\linewidth]{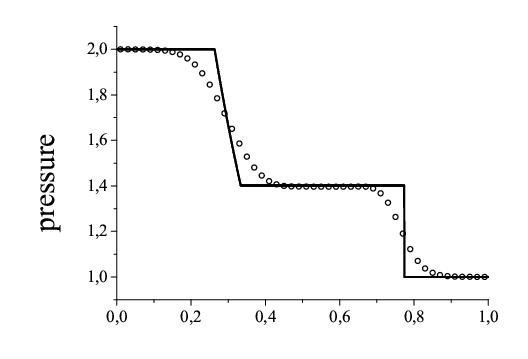} \\ [5mm]
\includegraphics[bb = 0 0 528 359, width=0.3\linewidth]{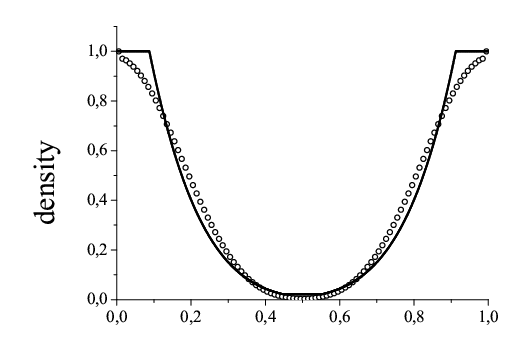}
\includegraphics[bb = 0 0 528 359, width=0.3\linewidth]{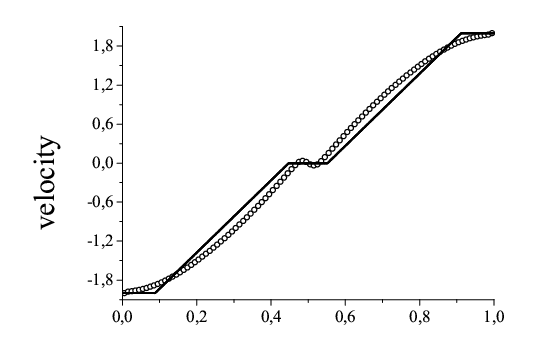}
\includegraphics[bb = 0 0 528 359, width=0.3\linewidth]{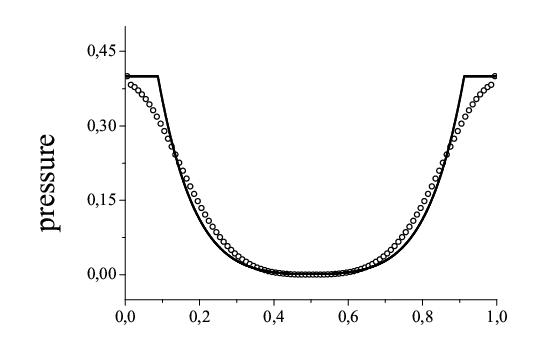} \\ [5mm]
\includegraphics[bb = 0 0 528 359, width=0.3\linewidth]{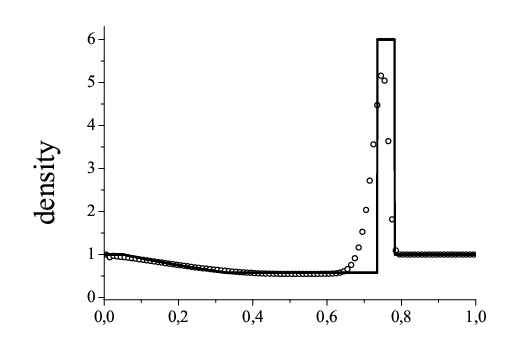}
\includegraphics[bb = 0 0 528 359, width=0.3\linewidth]{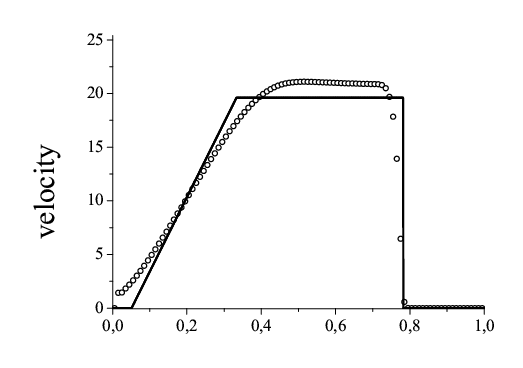}
\includegraphics[bb = 0 0 528 359, width=0.3\linewidth]{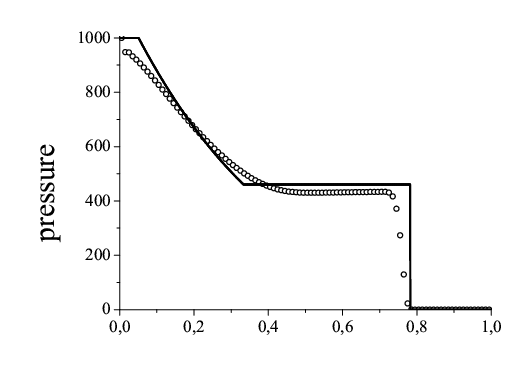}
\caption{Distribution of density, velocity and pressure in the simulation of the first test (top row)
the second test (middle row), the third test (bottom row). The solid line represents the exact solution, the dots indicate the result of the computation. The purpose of the first test is to determine the correctness of the
description of the contact discontinuity. In the second test, the gas with the same thermodynamic
parameters expands, producing a rarefaction region in the centre. The main goal of the third test is to check the
stability of the numerical method.}
\label{ShockTubeSimulation}
\end{figure}

Godunov tests are based on the on a shock wave simulation. 
Table \ref{ShockTube} shows the initial
configurations of shock tube for the tests.
The results of simulation are given in figures \ref{ShockTubeSimulation}.

\subsection{The Shock tube test for the Boltzmann moment equations}
\begin{figure}[ht]
\centering
\includegraphics[bb = 0 0 568 416, width=0.45\linewidth]{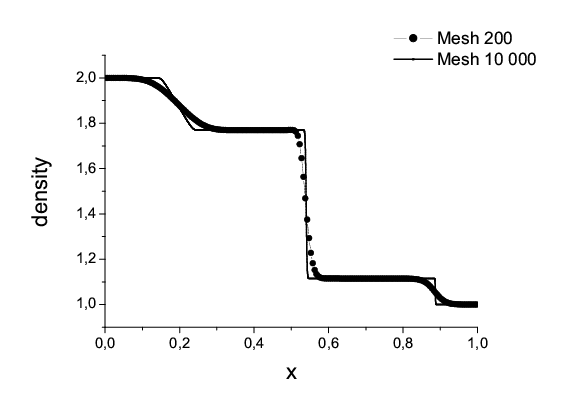}
\includegraphics[bb = 0 0 568 416, width=0.45\linewidth]{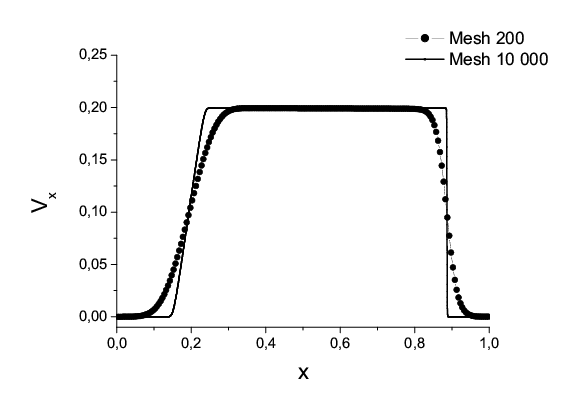} \\ [5mm]
\includegraphics[bb = 0 0 568 416, width=0.45\linewidth]{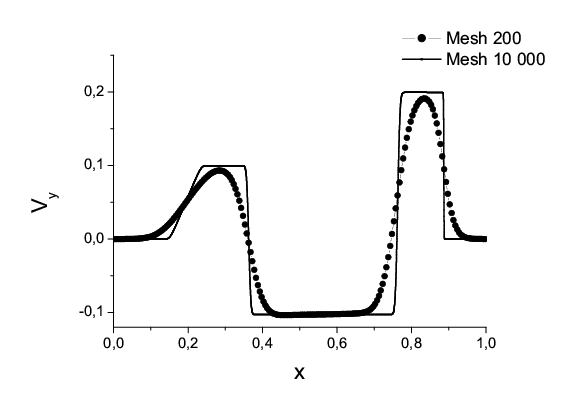} 
\includegraphics[bb = 0 0 568 416, width=0.45\linewidth]{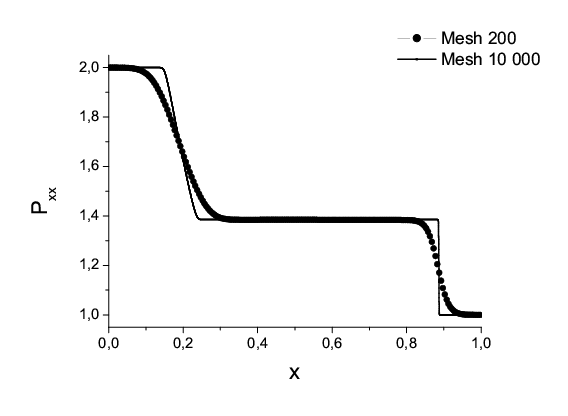} \\ [5mm]
\includegraphics[bb = 0 0 568 416, width=0.45\linewidth]{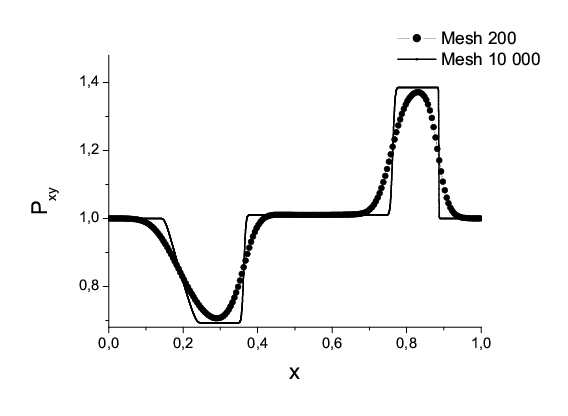}
\includegraphics[bb = 0 0 568 416, width=0.45\linewidth]{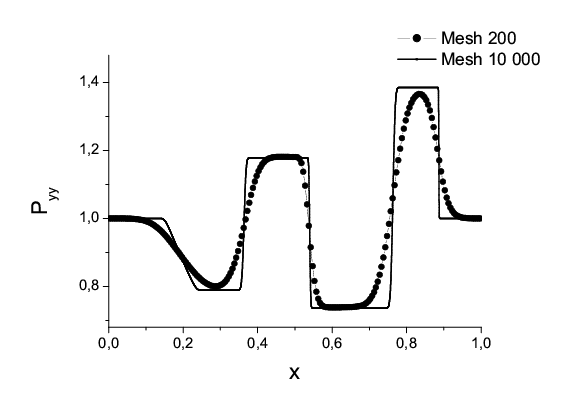} 
\caption{Distribution of density, velocity and the symmetric velocity dispersions tensor in the simulation 
of the Shock tube test for the Boltzmann moment equations. The solid line represents the exact solution, 
the dots indicate the result of the computation. The purpose of the test is to determine the correctness of the
description of the discontinuities.}
\label{ShockTubeBoltzmann}
\end{figure}

The initial configurations of the Shock tube test for the Boltzmann moment equations are
$$
\left[ \rho,\Pi_{xx},\Pi_{xy},\Pi_{xz},\Pi_{yy},\Pi_{yz},\Pi_{zz},u_{x},u_{y},u_{z} \right] = 
\left\{
\begin{array}{@{\,}r@{\quad}l@{}}
\left[ 2,2,1,1,1,1,1,0,0 \right], & x \le 0.5, \\ 
\left[ 1,1,1,1,1,1,1,0,0 \right], & x \le 0.5
\end{array} \right.
$$
The results of simulation are given in figures \ref{ShockTubeBoltzmann}.

\subsection{Aksenov test}
Consider the equations of one-dimensional gas dynamics in the dimensional form:
$$
\frac{\partial u}{\partial t} + u \frac{\partial u}{\partial x} = - \frac{1}{\rho} \frac{\partial p}{\partial x},
$$
$$
\frac{\partial \rho}{\partial t} + \frac{\partial \rho u}{\partial x} = 0,
$$
$$
\frac{p}{p_{0}} = \left ( \frac{\rho}{\rho_{0}} \right )^{\gamma},
$$
where $p$ - pressure, $\rho$ - density, $u$  - velocity, $\gamma$ - adiabatic index.

\begin{figure}[ht]
\centering
\includegraphics[bb = 0 0 514 359, width=0.45\linewidth]{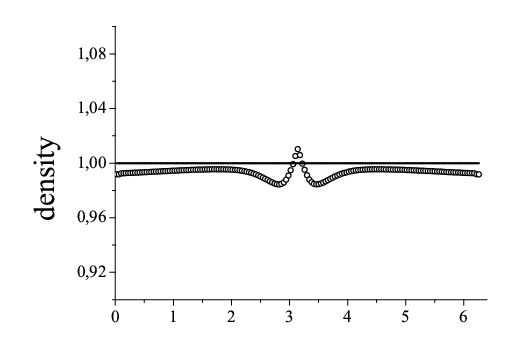}
\includegraphics[bb = 0 0 514 359, width=0.45\linewidth]{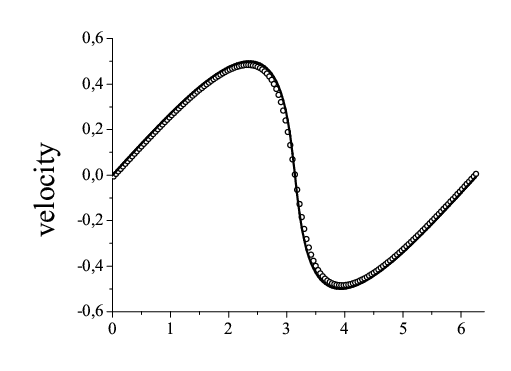}
\caption{Distribution of density, velocity in time $t = \pi/2$. The solid line shows exact solution, the points shows computation.}
\label{AksenovSimulation}
\end{figure}

As the characteristic variable to $l$ -- characteristic length, $\rho_{0}$ -- characteristic density, $p_{0}$ -- characteristic pressure.
Then the characteristic velocity is $u_{0} = \sqrt{\gamma p_{0} / \rho_{0}}$ and characteristic time is $t_{0} = l/\sqrt{\gamma p_{0} / \rho_{0}}$.
Selecting the dimensional quantities $l = 1$, $p_{0} = 1$, $\rho_{0} = 1$, $\gamma = 3$ and $\lambda = 1/(\gamma - 1)$,
$r = \rho^{1/2\lambda}$, $z = u / 2\lambda$.
Let us choose the initial data \citep{Aksenov_2001} $r = 1 + 0.5 cos(x)$, $z = 0$.
Then analytical solution for $[0;2\pi]$ is:
$$
r = 1 + 0.5 cos(x - zt) cos(rt),
$$
$$
z = 0.5 sin(x - zt) sin(rt).
$$
For time $t = \pi/2$ is $r(x) = 1$ and $z = 0.5 sin(x - zt)$.
The results of simulation are given in figures \ref{AksenovSimulation}.

Velocity is sufficiently well approximated numerical solution. Density has a jump in the center.
This jump is of the same principle as temperature jump in the third Godunov test.
Actually, we have to be something of type a trace of entropy in this test, which is formed as a result
of ''run-off'' gas in this region with zero velocity.
This feature focuses on a finite number of points and is reduced by splitting of mesh.

\subsection{Kelvin-Helmholtz and Rayleigh-Taylor Instability}
The gravitational instability is the basis for the physical formulation of the problem 
which leads to a mathematical incorrect. Numerical method must not suppress the physical instability. 
To check the correct reproduction of the instability a code has been
verificate on Rayleigh-Taylor and Kelvin-Helmholtz instability. Rayleigh-Taylor instability 
verifies playback capability of the gravitational term.
Kelvin-Helmholtz instability verifies playback capability of the nonlinear hydrodynamic turbulence.

The initial data for Rayleigh-Taylor instability: $[-0.5;0.5]^{2}$ -- domain, $\gamma = 1.4$ -- adiabatic index,
$$\rho_{0}(x) = \left\{
\begin{array}{@{\,}r@{\quad}l@{}}
1, & r \le 0, \\ 2, & r > 0
\end{array}\right.
$$
$p = 2.5 - \rho g y$ -- the hydrostatic equilibrium pressure, $g$ -- acceleration of free fall, $v_{y,0}(x,y) = A(y)[1+cos(2 \pi x)][1+cos(2 \pi y)]$,
where
$$A(y) = \left\{
\begin{array}{@{\,}r@{\quad}l@{}}
10^{-2}, & |y| \le 0.01, \\ 0, & y > 0.01
\end{array}\right.
$$
The initial data for Kelvin-Helmholtz instability: $[-0.5;0.5]^{2}$ -- domain, $\gamma = 1.4$ -- adiabatic index,
$$\rho_{0}(x) = \left\{
\begin{array}{@{\,}r@{\quad}l@{}}
1, & r \le 0, \\ 2, & r > 0
\end{array}\right.
$$
$$v_{x} = \left\{
\begin{array}{@{\,}r@{\quad}l@{}}
0.5, & |y| \le 0.25, \\ -0.5, & |y| > 0.25
\end{array}\right.
$$
$p = 2.5$ -- pressure, $v_{y,0}(x,y) = A(y)[1+cos(8 \pi x)][1+cos(8 \pi y)]$,
where
$$A(y) = \left\{
\begin{array}{@{\,}r@{\quad}l@{}}
10^{-2}, & ||y|-0.25| \le 0.01, \\ 0, & ||y|-0.25| > 0.01
\end{array}\right.
$$

\begin{figure}[ht]
\centering
\includegraphics[bb = 0 0 541 358, width=0.48\linewidth]{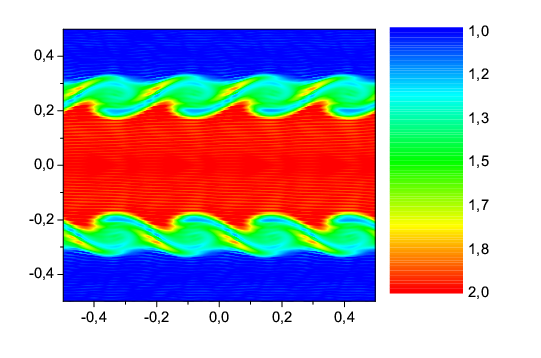}
\includegraphics[bb = 0 0 541 358, width=0.48\linewidth]{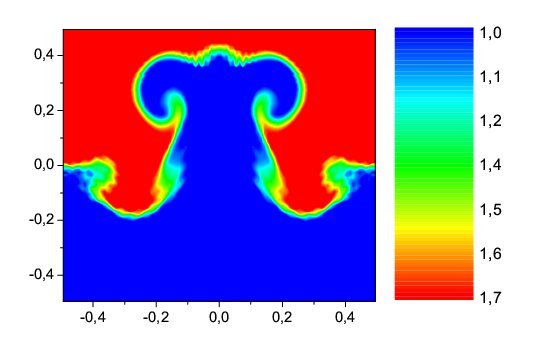}
\caption{Density in Kelvin-Helmholtz (left) and Rayleigh-Taylor (right) Instability simulation.}
\label{instability}
\end{figure}

The results of Kelvin-Helmholtz and Rayleigh-Taylor Instability simulation are given in figures \ref{instability}.

\subsection{Author's cloud collision test}
\begin{figure}[ht]
\centering
\includegraphics[bb = 0 0 541 358, width=0.48\linewidth]{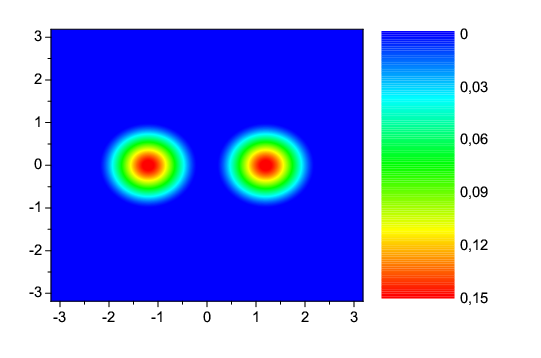}
\includegraphics[bb = 0 0 541 358, width=0.48\linewidth]{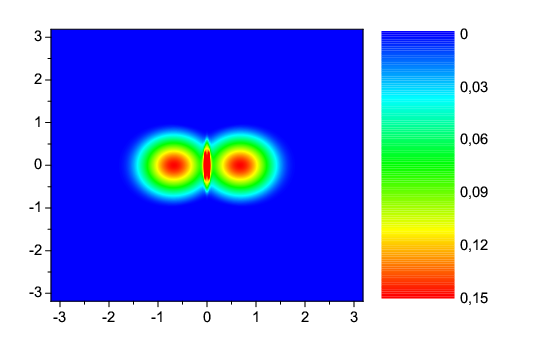} \\ [5mm]
\includegraphics[bb = 0 0 541 358, width=0.48\linewidth]{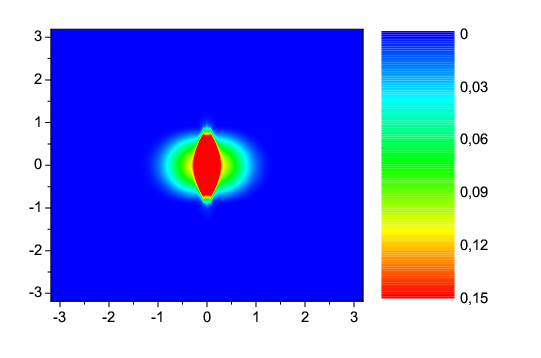}
\includegraphics[bb = 0 0 541 358, width=0.48\linewidth]{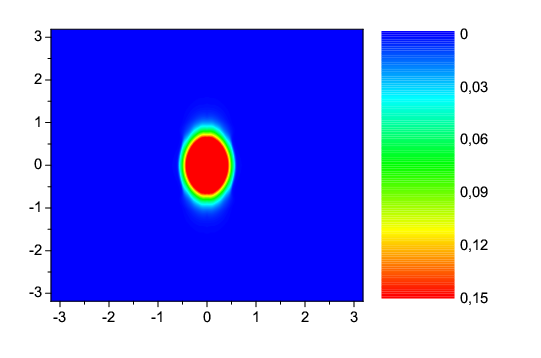} \\ [5mm]
\includegraphics[bb = 0 0 541 358, width=0.48\linewidth]{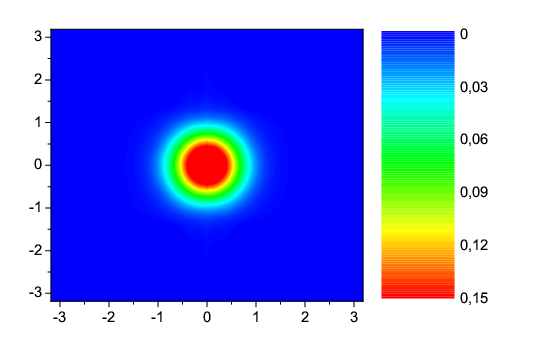}
\includegraphics[bb = 0 0 541 358, width=0.48\linewidth]{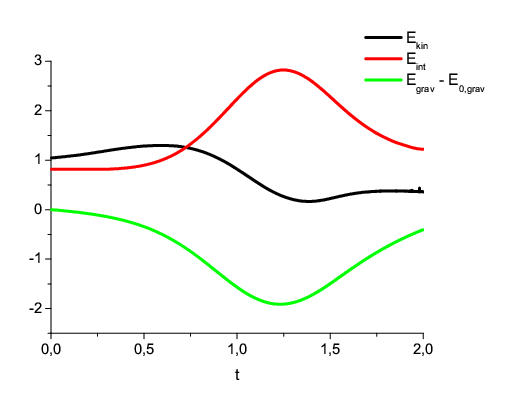} \\ [5mm]
\caption{Author's cloud collision test. Non-dimensional density distribution in dynamic. The behavior of different kinds of energy: inner energy (red line), kinetic energy (black line), potential energy (green line).
The mass of gas and total impulse are conserved because of the selected numerical scheme.
The conservation of energy was traced in the course of computation.}
\label{mywengen}
\end{figure}

Let us take the steady state gas sphere in the situation of hydrostatical equilibrium as the 
initial state for the gas dynamics equation. The density distribution is obtained from the 
gas dynamics equation system together with Poisson equation written in spherical coordinate:
$$
\frac{\partial p}{\partial r} = -\frac{M(r) \rho}{r^{2}}
$$
$$
\frac{\partial M}{\partial r} = 4 \pi r^{2} \rho
$$
$$
p = (\gamma - 1)\rho \epsilon
$$
The density distribution is:
$$
\rho_{0}(r) = \left\{
\begin{array}{@{\,}r@{\quad}l@{}}
1-r, & r \le 1, \\ 0, & r > 1.
\end{array}\right.
$$
Then pressure and gravitational potential is:
$$
p_{0}(r) = \left\{
\begin{array}{@{\,}r@{\quad}l@{}}
-\frac{\pi r^{2}}{36}(9r^{2}-28r+24)+\frac{5\pi}{36}, & r \le 1, \\ 0, & r > 1.
\end{array}\right.
$$
$$
\Phi_{0}(r) = \left\{
\begin{array}{@{\,}r@{\quad}l@{}}
-\frac{\pi}{3}(r^{3}-2r^{2})-\frac{2\pi}{3}, & r \le 1, \\ -\frac{\pi}{3r}, & r > 1.
\end{array}\right.
$$
The initial distance between two gas clouds $L_{0} = 2.4$, collision velocity $v_{collide} = 1.0$.
The results of simulation are given in figures \ref{mywengen}.

\subsection{Collapse of rotation cloud}
\begin{figure}[ht]
\centering
\includegraphics[bb = 0 0 516 418, width=0.48\linewidth]{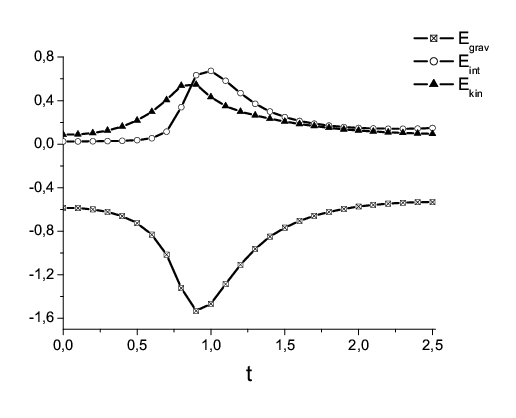}
\caption{Behavior of various types of energy in the collapse of molecular clouds simulation}
\label{collapse}
\end{figure}

For research the possibility of modeling the collapse of rotating molecular clouds we simulate the gas cloud bounded by the sphere with radius -- ${R_0} = 100$ pc, mass of cloud -- ${M_g} = {10^7}{M_ \odot }$, 
density -- $\rho \left( r \right) \simeq {1/r}$, temperature -- $T \approx 2000$ K, angular velocity -- $\omega  = 21$ km/sec, adiabatic index -- $\gamma = 5/3$, speed of sound -- $c \approx 3.8$ km/sec.
The main characteristic parameters are: ${L_0} = 100$  pc,   ${\rho _0} = 1.2 \cdot {10^{ - 18}}$ kg/m$^{3}$, ${v_0} = 21$  km/sec. Then Non-dimensional density -- $\rho  = 1.0$, pressure -- $p = 2 \times {10^{ - 2}}$, angular velocity -- $\omega  = 1$,
adiabatic index -- $\gamma = 5/3$, domain -- ${\left[ {0;6.4} \right]^3}$.
In this research the behavior of energy has quantitatively \ref{collapse} coincided with the results of other authors \citep{Berczik_2005} .

\section{Numerical simulation of a galaxies collision}
\begin{figure}[ht]
\centering
\includegraphics[bb = 0 0 541 358, width=0.3\linewidth]{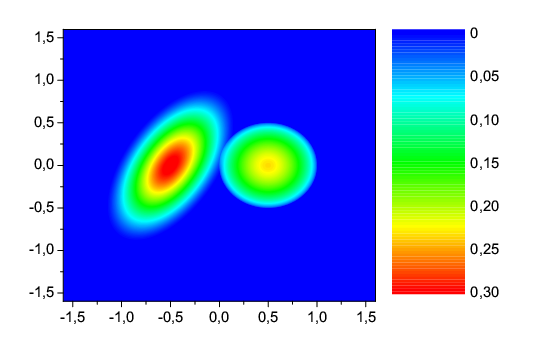}
\includegraphics[bb = 0 0 541 358, width=0.3\linewidth]{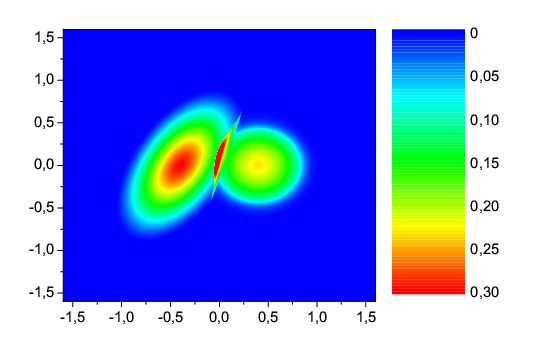}
\includegraphics[bb = 0 0 541 358, width=0.3\linewidth]{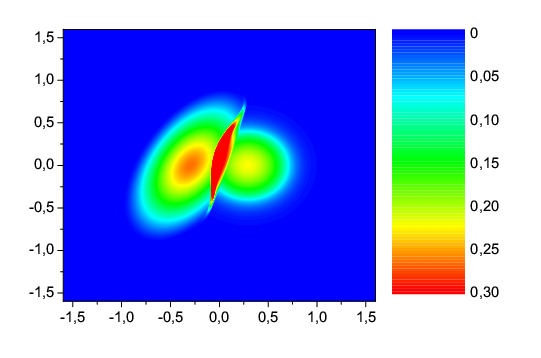}
\caption{Non-dimensional density of gas clouds at times: the initial configuration (left),
$1 \cdot 10^{14}$ sec (middle), $2 \cdot 10^{14}$ sec (right)}
\label{supersimulation}
\end{figure}

The main purpose of code GPUPEGAS is modeling the galaxies collision of different types and at different angles.
As a model problem, we consider the collision of disk galaxies at an angle.
The first cloud is given as spherical domain filled uniformly with gas $M_{gas} = 16 \cdot 10^{41}$ kg,
The second cloud is given in respect of the ellipsoid axes 1:2:1, inclined at 45 degrees to the axis of the collision.
The clouds move in the opposite directions with the velocities  $v_{cr} = 600$ km/sec.
The figure \ref{supersimulation} shows the evolution of the collision and ''slim'' splash at a interacting galaxies.
The calculation was made using 96 GPU-accelerators cluster NKS-G6 of Siberian Supercomputer Center ICMMG SB RAS on the mesh $1024^{3}$ and $10^{5}$ time steps.

\subsection{The passage scenario of a central collision of two galaxies}
\begin{figure}[ht]
\centering
\includegraphics[bb = 0 0 541 358, width=0.45\linewidth]{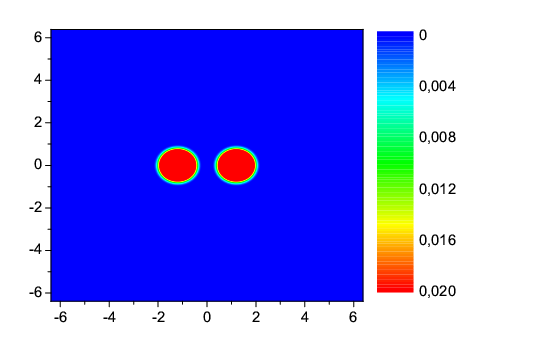}
\includegraphics[bb = 0 0 541 358, width=0.45\linewidth]{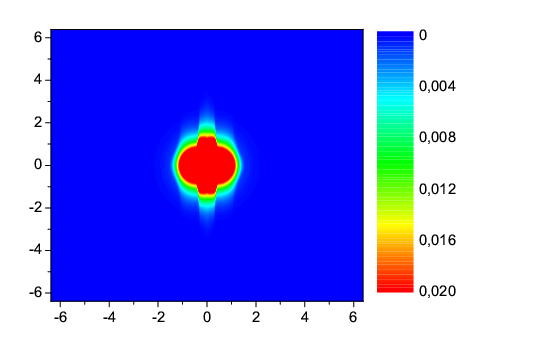} \\ [5mm]
\includegraphics[bb = 0 0 541 358, width=0.45\linewidth]{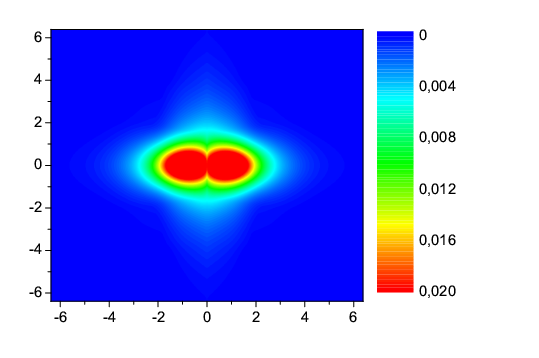}
\includegraphics[bb = 0 0 541 358, width=0.45\linewidth]{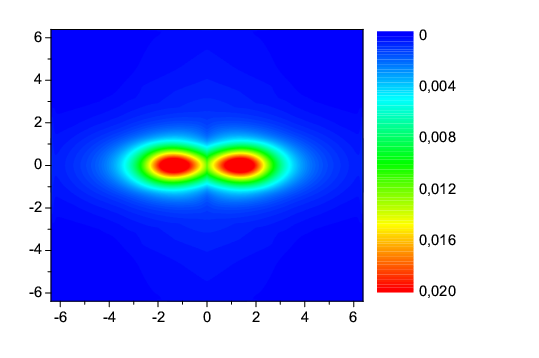}
\caption{Non-dimensional gas density in the collision plane, the scenario with the passage of galaxies.
The moments of time $1 \cdot 10^{14}$ sec -- collision begins (top left), $10^{15}$ sec -- the moment of collision (top right)
$2.5 \cdot 10^{15}$ sec -- start of the passage of galaxies (bottom left),
$4.6 \cdot 10^{15}$ sec -- final of the passage of galaxies (bottom right)}
\label{twophase}
\end{figure}

We should show the possibility of scenarios galaxies passage in the two-phase model \citep{Vshivkov_2011}.
Two self-gravitating gas clouds are set in the 3D computational domain at the initial moment.
Both clouds have the same distributions of gas parameters. Each cloud is a spherical domain filled uniformly 
by gas with the mass $M_{gas} = 16 \cdot 10^{41}$ kg and the stars and dark matter with the mass $M_{gas} = 16 \cdot 10^{41}$ kg.
The clouds move in the opposite directions with the velocities  $v_{cr} = 800$ km/sec.
We should repeat the passage scenario of a central collision of two galaxies in two-phase model.
The figure \ref{twophase} shows the evolution of the passage scenario of a central collision of two galaxies.

\section{Conclusions and future work}
A new Hydrodinamical code GPUPEGAS is described for simulation of interacting galaxies on a hybrid supercomputer by means GPU.
The code is based on combination of Godunov method as well as on the original implementation of FlIC method, specially adapted for GPU-implementation. Fast Fourier Transform is used for Poisson equation solution in GPUPEGAS.
Software implementation of the above methods was tested on classical gas dynamics problems, new Aksenov's test and 
classical gravitational gas dynamics problems. The Boltzmann moment equations approach was used for modelling of stars and dark matter. The possibility of scenarios galaxies passage in the two-phase model is shown.
The scalability of GPUPEGAS computational accelerators is shown. Maximum speed-up factors of 55 (as compared with 12.19 on GAMER code \citep{Schive_2010}) are demonstrated using one GPU. Maximum efficiency 96 \% (as compared with 65 \% on GAMER code on 16GPUs \citep{Schive_2010}) are demonstrated using 60 GPUs cluster NKS-G6 of Siberian Supercomputer Center ICMMG SB RAS. 

\acknowledgments
The research work was supported by the Federal Program ''Scientific and scientific-pedagogical cadres innovation Russia for 2009-2013'' and Federal Programme for the Development of Priority Areas of Russian Scientific \& Technological Complex 2007-2013,
Russian Ministry Education and Science, RFBR grant 12-01-31352 for junior researchers and by the Grant of
the President of Russian Federation for the support of young scientists number MK -- 4183.2013.9.

\begin{deluxetable}{cccccc}
\rotate
\tablecaption{The main properties of the widely used software packages}
\tablewidth{0pt}
\tablehead{
\colhead{Code} & \colhead{Hydrodynamical method} & \colhead{Poisson solver} & \colhead{HPC technologies} & \colhead{Correctness checking} }
\startdata
  Hydra       & SPH & Adaptive P$^{3}M + FFT$ & High Performance Fortran & +  \\
  Gasoline    & SPH & Tree code + Multipole Method & MPI & -- \\
  GrapeSPH    & SPH & Direct Summation  & GRAPE & -- \\
  GADGET-2    & SPH & TreePM + FFT & MPI & -- \\
  NIRVANA     & AMR+HLL & Multigrid & MPI & -- \\
  FLASH       & AMR+PPM & Multigrid & MPI & + \\
  ZEUS-MP     & Finite difference method & FFT+Multigrid & MPI & -- \\
  ENZO        & AMR+PPM & FFT+Multigrid & MPI & -- \\
  RAMSES      & AMR+HLLC & Multigrid+CG & OpenMP+MPI & + \\
  ART         & AMR+MUSCL & FFT & MPI & -- \\
  Athena      & Roe’s solver & FFT & MPI & -- \\
  Pencil Code & Finite difference method & FFT & HPF+MPI & + \\
  Heracles    & MUSCL & CG & MPI & -- \\
  Orion       & AMR+MUSCL & Multigrid & -- & + \\
  Pluto       & AMR+HLLC & Analytical & MPI & -- \\
  CASTRO      & AMR+PPM & Multigrid & MPI+OpenMP &  --\\
  GAMER       & AMR+TVD & FFT+SOR & CUDA+MPI & -- \\
  BETHE-Hydro & Arbitrary Lagrangian-Eulerian & Matrix Inverse & -- & -- \\
  AREPO       & Moving mesh + MUSCL & TreePM + FFT & MPI & -- \\
  CHIMERA     & Moving mesh + PPM & Analytical & -- & -- \\
  PEGAS       & FlIC+Godunov & FFT & MPI & + \\
\enddata
\label{PropTable}
\end{deluxetable}

\begin{deluxetable}{ccccc}
\tablecaption{The initial state of the shock tube.}
\tablewidth{0pt}
\tablehead{ \colhead{Parameter} & \colhead{Test 1} & \colhead{Test 2} & \colhead{Test 3} }
\startdata
  $\rho_{L}$ & 2 & 1 & 1 \\
  $v_{L}$ & 0 & -2 & 0  \\
  $p_{L}$ & 2 & 0.4 & 1000  \\
  $\rho_{R}$ & 1 & 1 & 1 \\
  $v_{R}$ & 0 & 2 & 0 \\
  $p_{R}$ & 1 & 0.4 & 0.01 \\
  $x_{0}$ & 0.5 & 0.5 & 0.5 \\
  $t$ & 0.2 & 0.15 & 0.012 \\
\enddata
\label{ShockTube}
\end{deluxetable}


\begin{thebibliography}{}
\bibitem[Tutukov \& Fedorova(2006)]{Tutukov_2006}Tutukov, A., Fedorova, A. 2006, Astron. Rep., 50, 785
\bibitem[Tutukov et al.(2011)]{Tutukov_2011}Tutukov, A., Lazareva, G., Kulikov, I. 2011, Astron. Rep., 55, 770
\bibitem[Gingold \& Monaghan(1977)]{Gingold_1977}Gingold, R.A., Monaghan, J.J. 1977, MNRAS, 181, 375
\bibitem[Luci(1977)]{Luci_1977}Luci, L.B. 1977, ApJ, 82, 1013
\bibitem[Collela \& Woodward(1984)]{Collela_1984}Collela, P., Woodward, P.R. 1984, J. Comp. Phys., 54, 174
\bibitem[O’Shea et al.(2005)]{Norman_2005}O’Shea, B., Bryan, G., Bordner, J., Norman, M., Abel, T., Harkness, R., \& Kritsuk, A. 2005, Lect. Notes Comput. Sci. Eng., 41, 341
\bibitem[Hockney \& Eastwood(1981)] {Hockney_1981}Hockney, R.W., \& Eastwood, J.W. 1981, Computer Simulation Using Particles (New York: McGraw-Hill)
\bibitem[Couchman(1991)]{Couchman_1991}Couchman, H. M. P. 1991, ApJ, 368, L23
\bibitem[Barnes \& Hut(1986)]{Barnes_1986}Barnes, J., \& Hut, P. 1986, Nature, 324, 446
\bibitem[Dubinski et al.(2004)]{Dubinski_2004}Dubinski, J., Kim, J., Park, C., \& Humble, R. 2004, New Astron., 9, 111
\bibitem[Fedorenko(1961)]{Fedorenko_1961}Fedorenko, R. 1961, USSR Comput. Math. \& Math. Phys., 1, 1092
\bibitem[Godunov(1959)]{Godunov_1959}Godunov, S. K. 1959, Math. Sb., 47, 271
\bibitem[Kulikovsky et al.(2001)]{Kulikovsky_2001}Kulikovskii, A. G., Pogorelov, N. V., \& Semenov, A. Yu. 2001, Mathematical Aspects of Numerical Solution of Hyperbolic Systems (in Russian;Moscow:Fizmatlit)
\bibitem[Toro(1999)]{Toro_1999}Toro, E.F. 1999, Riemann Solvers and Numerical Methods for Fluid Dynamics (Heidelberg:Springer-Verlag)
\bibitem[Courant et al.(1952)]{Courant_1952}Courant, R., Isaacson, E., Rees, M. 1952, Communications on Pure and Applied Mathematics, 5, 243
\bibitem[Roe(1997)]{Roe_1997}Roe, P. 1997, J. Comp. Phys., 135, 250
\bibitem[Engquist \& Osher(1981)]{Engquist_1981}Engquist, B., Osher, S.J. 1981, Math. Comp., 36, 321
\bibitem[Harten et al.(1983)]{Harten_1983}Harten, A., Lax, P.D., Van Leer, B. 1983, Soc. Indust. \& Appl. Math. Rev., 25, 35
\bibitem[Einfeld(1988)]{Einfeld_1988}Einfeld, B. 1988, Soc. Indust. \& Appl. Math. J. Num. Analys., 25, 294
\bibitem[Batten et al.(1997)]{Batten_1997}Batten, P., Clarke, N., Lambert, C., \& Causon, D.M. 1997, Soc. Indust. \& Appl. Math. J. Comp., 18, 1553
\bibitem[Van Leer(1979)]{VanLeer_1979}Van Leer, B. 1979, J. Comput. Phys., 32, 101
\bibitem[Pearcea \& Couchman(1997)]{Couchman_9603116}Pearcea, F.R., Couchman, H.M.P. 1997, New Astronomy, 2, 411
\bibitem[Wadsley et al.(2004)]{Wadsley_2004}Wadsley, J.W., Stadel, J., Quinn, T. 2004, New Astronomy, 9, 137
\bibitem[Matthias(1996)]{Matthias_1996}Matthias, S. 1996, MNRAS, 278, 1005
\bibitem[Springel(2005)]{Springel_2005}Springel, V. 2005, MNRAS, 364, 1105
\bibitem[Jin \& Xin(1995)]{Jin_1995}Jin, S., Xin, Z. 1995, Commun. Pure Appl. Math., 48, 235
\bibitem[Godunov et al.(2011)]{Godunov_2011}Godunov, S.K., Manuzina, Yu.D., Nazareva, M.A. 2011, Comp. Math. Comp. Phys, 51, 88
\bibitem[Ziegler(2005)]{Ziegler_2005}Ziegler, U. 2005, A\&A, 435, 385
\bibitem[Mignone et al.(2005)]{Mignone_2005}Mignone, A., Plewa, T., Bodo, G. 2005, ApJ, 160, 99
\bibitem[Norman et al.(2006)]{Hayes_2006}Hayes, J., Norman, M., Fiedler, R. et al. 2006, ApJS, 165, 188
\bibitem[Teyssier(2002)]{Teyssier_2002}Teyssier, R. 2002, A\&A, 385, 337
\bibitem[Kravtsov et al.(2002)]{Kravtsov_2002}Kravtsov, A., Klypin, A., Hoffman, Y. 2002, ApJ, 571, 563
\bibitem[Stone et al.(2008)]{Stone_2008}Stone, J. et al. 2008, ApJS, 178, 137
\bibitem[Brandenburg \& Dobler(2002)]{Brandenburg_0109497}Brandenburg, A., Dobler, W. 2002, Comp. Phys. Comm., 147, 471
\bibitem[Schive et al.(2010)]{Schive_2010}Schive, H., Tsai, Y., Chiueh, T. 2010, ApJ, 186 457
\bibitem[Murphy \& Burrows(2008)]{Murphy_2008}Murphy, J., Burrows, A. 2008, ApJS, 179, 209
\bibitem[Springel(2009)]{Springel_2009}Springel, V. 2010, MNRAS, 401, 791
\bibitem[Bruenn et al.(2009)]{Bruenn_2009}Bruenn, S. et al. 2009, J. Phys., 180, 393
\bibitem[Vshivkov et al.(2011,1)]{Vshivkov_2011}Vshivkov, V., Lazareva, G., Snytnikov, A., Kulikov, I., \& Tutukov, A. 2011, ApJS, 194, 47
\bibitem[Gonzalez et al.(2007)]{Gonzalez_2007}Gonzalez, M., Audit, E., Huynh P. 2007, A\&A, 464, 429
\bibitem[Krumholz et al.(2007)]{Krumholz_2007}Krumholz, M.R., Klein, R.I., McKee, C.F., \& Bolstad, J. 2007, ApJ, 667, 626
\bibitem[Mignone et al.(2007)]{Mignone_2007}Mignone, A. et al. 2007, ApJS, 170, 228
\bibitem[Almgren et al.(2010)]{Almgren_2010} Almgren, A. et al. 2010, ApJ, 715, 1221
\bibitem[Feng et al.(2011)]{Feng_2011}Feng, Y. et al., 2011, ApJS, 197, 18
\bibitem[Ferrari(2010)]{Ferrari_2010}Ferrari, A. 2010, HPC on vector systems 2009, 4, 179
\bibitem[Straalen(2009)]{Straalen_2009}Van Straalen, B. 2009, Proc. IPDPS 09, 1
\bibitem[Mitchell et al.(2012)]{Mitchell_2012}Mitchell, N., Vorobyov, E., Hensler, G. 2012, MNRAS, 428, 2674
\bibitem[Sutherland \& Dopita(1993)]{Sutherland_1993}Sutherland, R. S., \& Dopita, M. A. 1993, ApJS, 88, 253
\bibitem[Vshivkov et al.(2006)]{Vshivkov_2006}Vshivkov, V. A., Lazareva, G. G., \& Kulikov, I. M. 2006, Comp. Tech., 11, 27 (in Russian)
\bibitem[Vshivkov et al.(2007)]{Vshivkov_2007}Vshivkov, V., Lazareva, G., \& Kulikov, I. 2007, Opt. Instrum. Data Proc., 43, 530
\bibitem[Vshivkov et al.(2011,2)]{Vshivkov_2011_2}Vshivkov, V., Lazareva, G., Snytnikov, A., Kulikov, I., \& Tutukov, A. 2011, J. Inv. Ill-pos. Prob., 19, 1, 151
\bibitem[Aksenov(2001)]{Aksenov_2001}Aksenov, A.V. 2001, Doklady Akademii Nauk, 381, 2, 176 (in Russian)
\bibitem[Petrov \& Berczik(2005)]{Berczik_2005}Petrov, M.I., Berczik, P.P. 2005, Astronomische Nachrichten, 326, 505
\end{thebibliography}
\end{document}